\documentclass[sn-apa]{sn-jnl}

\usepackage{graphicx}%
\usepackage{multirow}%
\usepackage{amsmath,amssymb,amsfonts}%
\usepackage{amsthm}%
\usepackage{mathrsfs}%
\usepackage[title]{appendix}%
\usepackage{xcolor}%
\usepackage{textcomp}%
\usepackage{manyfoot}%
\usepackage{booktabs}%
\usepackage{algorithm}%
\usepackage{algorithmicx}%
\usepackage{algpseudocode}%
\usepackage{listings}%

\theoremstyle{thmstyleone}%
\theoremstyle{thmstyletwo}%

\theoremstyle{thmstylethree}%

\begin{document}

\title[AI Gender Bias, Disparities, and Fairness: Does Training Data Matter?]{AI Gender Bias, Disparities, and Fairness: Does Training Data Matter?}


\author*[1]{\fnm{Ehsan} \sur{Latif}}\email{ehsan.latif@uga.edu}

\author[2]{\fnm{Xiaoming} \sur{Zhai}}\email{xiaoming.zhai@uga.edu}
\author[3]{\fnm{Lei} \sur{Liu}}\email{lliu001@ets.org}


\affil*[1,2]{\orgdiv{AI4STEM Education Center}, \orgname{University of Georgia}, \orgaddress{ \city{Athens}, \postcode{30605}, \state{Athens}, \country{USA}}}

\affil[3]{ \orgname{Educational Testing Service}, \orgaddress{\city{Princeton}, \state{New Jersey}, \country{USA}}}



\abstract{This paper examines one of the pressing concerns of gender biases in the human subject data used to train large language models. As a part of this discussion it summarizes the findings about gender bias in automatic scoring of student-written responses. Therefore, the first research question is to analyze the potential gender bias, gender inequality, and gender equity in usually gender-averaged AI training samples (i.e., male, female, and a combination of both). Using the fine-tuned version of the BERT model and combined with GPT-3.5, this research employs more than 6000 human-graded student responses from seventy males and seventy females across six tasks. The study employs three distinct techniques for bias analysis: The metrics used in this work are scoring accuracy difference to assess the degree of bias, mean score gaps by gender (MSG) to determine gender disparity, and Equalized Odds (EO) to measure fairness. This has been found to be true concerning mixed-gender trained models, in which results have said the scoring accuracy difference between male- and female-trained models is insignificant, therefore supporting that no significant scoring bias exists. In agreement with BERT and GPT-3.5 was that mixed gender trained models produced fewer MSG and non-disparate predictions than the humans are capable of, while the gender-specifically trained models produced a higher MSG as compared to the humans, meaning that unbalanced training data in the generation of algorithmic models, only widens the existing gender gaps. The study conducted under the EO analysis indicates that mixed-gender trained models helped to produce a better outcome than gender-specific training. Altogether, the research leads to the conclusion that gender-sparse data do not produce scoring prejudice in the sex-based sense but contribute to the expansion of gender gaps as well as diminished scoring equity.}

\maketitle


\section{Introduction}
Recent AI advancements have reached the core of human existence indicators, disturbing practices, and industries. However, that is possible only in theory because when using people to perform the job of AI, they will bring inbuilt biases, especially gender biases, which are always a sure recipe for disaster. Such biases are extended in universally proper word embeddings, and supervised machine learning categorizations are emphasized \citep{bolukbasi2016man,hall2023systematic}. Discrimination against women in AI automatic scoring systems automatically prop up socio-cultural prejudices and inequalities \citep{zhai2023ai}. Consequently, \cite{cirillo2020sex, leavy2018gender,nadeem2020gender} have illustrated how diversity affects machine learning decisions out loud these questions of bias.

However, this belief can also create high social costs when bias is inflated, especially pseudo-AI biases that widen the gap between the technology-disadvantaged groups, which include females. Some of the potential threats that may be found in the new generation of AI include pseudo-AI bias, which is already a known term defined as a false assumption of AI bias \citep{zhai2022pseudo}. For instance, if the AI scoring leads in the way that particular differences between male and female students are found, but these differences are not systematic. Such a result will not be a bias but rather a mistake. People often explain such an error as 'bias' that originates from AI, which is one of the manifestations of pseudo-AI bias. Vague assertions that there is a pseudo-AI bias can also amplify an unjustified fear about AI. Therefore, detecting AI gender bias is crucial to avoid the confusion of pseudo-AI bias with genuine AI bias.

This research work is beneficial in providing a certain kind of value to the existing literature that discusses gender bias with AI, as lots of work within this area of study has recently been pointed out to lack adequate data. As equally relevant to bureaucratic enhancement as not removing gender discrimination from technology, \cite{franzoni2023gender} notes that innovation continues in gender discrimination in the promotions of the police force. Still, Franzoni scarcely discussed the socio-cultural bias of AI and its potential consequences, thus not having a principal emphasis on the technical side. From this view, a detailed iterative literature review of socio-technical gender bias in AI algorithms, as composed by \cite{hall2023systematic}, contributes to that activity to point out the fact that when a given individual does not have a sure other form of bias that is not current in the AI, he or she can end up affected a lot by that bias and vice versa; thus, it requires a realistic and tangible means of finding that bias even though their work has no empirical context in terms of comparing specific ML models or datasets for training those models, analysis of the bias was conducted and, therefore, the existing literature informed theoretical framework of the study. Furthermore, \cite{lima2023gender} state that the current AI systems are gender biased, working sensitively to gender approaches, mentioned that they need to be further polished, and strategies need to be debated about how to train an AI. However, they did not discuss such an important question: what if the data sets used are imbalanced, and this AI reflects such imbalances? To the above question, this research fills the gap through the survey method combined with Econometric analysis.

This study also satisfies the following goals independently: analyzing gender bias through extensive studies and modified algorithms. \cite{li2022using}, in their study, did not expound on fair AI in outcome predictions, as pointed out above, nor did they use mixed gender in moderation of bias. Therefore, \cite{lu2020gender} offer insights into where the bias may stem from and what may be done to remove or mitigate it; however, they fail to explore the impact of training using mixed and gender-discriminating data sets. Briefly, \cite{sun2019mitigating} outlined the strategies used in the literature to mitigate gender bias in natural language processing and the approaches to tackle it while avoiding datasets with both genders. This paper aims to investigate the origins of the emerging gender bias that is forecasted through the utilization of artificial intelligence and the relationship that it might have with biased training sets.

As the Ethical Evaluation Framework provided by \cite{slimi2023navigating} and the Advantages listed by \cite{o2023gender} indicate, this paper deals with the ethical issue of fairness and equity in AI systems. Such is the case in \cite{slimi2023navigating}. From this analysis, it can be concluded that natural AI systems that reflect the current bias are inadequate. There is a need to design systems that do not reflect the current bias, primarily when AI technology will be used in learning sectors that support equality. On the same note, \cite{o2023gender} analyzes the reinforcement and reduction of gender bias in AI technologies, barriers, and possibilities, which offer insights on increasing chances of creating AI with less bias. Their work implies the complexity of eradicating sexist AI by giving a clear account of the process that underpins the creation of non-sexist training data, which is highly technical and socially layered. However, there is scarce knowledge on how to increase fairness measures and prevent bias from using training data containing skewed gender data.

This research contributes to the field by providing empirical evidence that targeted mix-gender model training can reduce gender bias. \cite{lima2023gender} offer a guide to the emancipative design of AI in education and other arenas. This work contributes to filling the identified gap in the literature by concentrating on how the training data affects bias. \cite{manresa2021assessing}) discuss the application of explainable AI in evaluating gender bias in AI models and the significance of the approaches in reducing bias. \cite{nemani2023gender} undertook research focusing on distinguishing various techniques concerning bias and the capacity of transformer models to handle the bias. While the data collected from their survey is advantageous in today's Gender bias literature, similar to the previous surveys, the authors' new assertion lacks empirical evidence that compares mixed-gender versus gender-specific training data. This study builds on their work to systematically test in an empirical study how changes in the structure of the training data utilized by the machine learning algorithms under investigation influence gender bias, disparity, and fairness in the final scores allocated by AI.

To address the research gaps in previously discussed studies, we extend the work reported by \cite{slimi2023navigating} and \cite{o2023gender} by exploring the applicability of the proposed mixed-gender data approach in the male and female training data samples. In addition, this study aligns with the elements of ethics highlighted by \cite{manresa2021assessing} and \cite{nemani2023gender} to ensure gender bias issues are addressed using a more complex approach. By extension, this means that while mixed-gender training data do not create scoring bias, they greatly assist in the elimination of gender bias and the enhancement of the anti-bias quotient of AI measures, hence aiding in the fabrication of ethical AI systems that are for the welfare of society.

As for broader patterns, one has to refer to how fine-tuned versions of BERT and GPT-3.5 affect the dissemination of information from broad perspectives. The study describes how bias, unfair treatment, and inconsistent scoring by the machine may be detected by comparing the student responses graded by the human experts. The study also addresses the gaps in the literature by assessing the mixed-trained model relative to gender-separate trained models. It demonstrates that mixed-trained models have an advantage in attaining a reduced mean score gap and generating more equal results. The study suggests that while training the AI model on biased data does not result in the creation of bias, the performance improves gender characteristics to influence fairness, an area of study many previous studies failed to explore. This work focuses on addressing the following research questions:
\begin{enumerate}
    \item How do gender-unbalanced training samples contribute to gender bias in automatic scoring?
    \item How do gender-unbalanced training samples affect AI scoring disparity by gender?
    \item How do gender-unbalanced training samples contribute to AI gender fairness for automatic scoring?
\end{enumerate}

\section{Background}
\subsection{Automatic Scoring in Education}
Today's classroom assessments are not like those of the past because of the availability of AI. As \cite{gonzalez2021artificial} have revealed in their recent comprehensive and highly detailed systematic review of the studies concerning the use of AI in developing student assessments, some methodologies have been employed and for which the countries of origin reported information concerning their impact. \cite{zhai2021practices} are the authors of an article discussing the state of the art of and practice regarding AI-based assessments in science concerning the development of their latest work where they present their findings. It reveals that many kinds of AI, such as machine learning and natural language processing, have been justified in that they can correctly assess the students' answers in all subjects.

Nevertheless, the formation of automatic scoring is not an easy task. \cite{zhai2022applying} build a study on applying machine learning for science assessments, which helps to understand how hard it is to assess student models correctly. Automatic scoring requires accuracy; therefore, the measures include agreed scores between the machines and humans. \cite{zhai2021meta} recently published a meta-analysis of machine learning-based science assessments, and the results showed partial consistency in human-machine agreement. It also expands on why agreements between machine and human scores are low. Their conclusions suggest that while, in some ways, AI can achieve a level of similarity to the humanly scored version, there are nonetheless disparities because of intra (such construct as construct complexity) and extra AI (such as algorithms) factors to their automatic assessments. It is still necessary to address the task or permanently update the assessment of individual algorithm parameters within educational environments.

Advancements in AI have not only made the automatic scoring process fast but also made it personalized and adaptive to enhance the student learning experience \citep{lee2023applying, zhai2020applying}. Researchers have used these models in automatically grading science writing \citep{latif2023artificial, lee2023multimodality}. This study focuses on the AI bias for automatic scoring of written responses of different genders; hence, all the defined terminologies revolve around the deviation of automatic scoring to analyze AI bias. Studies also highlight the ethical concerns of the multifaceted implementation of AI in education and suggest keeping the social and moral values along with technological advancements \citep{berendt2020ai}. \cite{felix2020role} and \cite{qadir2023engineering} share their perspectives on the fitness of AI in education, with changing applications of AI from fortifying teaching methodologies to assisting administrative work. This study addresses these ethical concerns by evaluating the impact of mixed dataset (containing responses from both genders of students) training of machine learning models for automatic scoring on AI bias.

\subsection{AI Gender Bias, Disparities, and Fairness}
Gender bias, disparities, and fairness – in AI, particularly educationally, as we find ourselves – are topics that remain surrounded by half-truths and facts. Subsequently, we define bias for AI systematically like \cite{larrazabal2020gender} since bias entails systematic rather than random error in which we study the calculated error in the scoring accuracy in the trained model of AI. For instance, if the machines award one gender a better score than the other, this can trigger debates on AI partiality to the said gender. Gender disparities in AI are defined as mean scoring gaps in automatically predicted machine scores for responses written by students of different genders; this conceptualization is based on the \cite{organisation2018bridging} definition. For instance, high variability of average scores between genders can increase gender inequality in AI applications.

Moreover, AI fairness as equity \citep{holstein2019fairness} is also discussed; in other words, if the normalized gender responses received scores close to one another by machines, then this represents equity in AI. The AI model is hypothesized to be bias-free if the ratio of False Positive to True Positive of automatically generated scores for the responses written by different genders is equal. This section aims to flesh out these aspects to gain a better knowledge and perception of how and to what extent gender issues are present in AI systems.

Lack of context, lack of related knowledge, and jumping to the wrong conclusions are invariably present at the heart of most misconceptions regarding the gender bias of AI. One such myth is that AI is not biased since it uses Algorithms and Data. However, as stated in \cite{bolukbasi2016man} and \cite{hardt2016equality}, biases are not only in the training data used in the designing of the algorithms but also in its operations; they may appear to form words or phrases that are neutral with regards to gender. The next myth is that the issue of gender AI bias is the problem at the technical level only. To solve these problems, it is necessary to collect more data and at least create better algorithms. These are basic ingredients that contribute towards the success of any AI solution, as noted by \cite{lu2020gender} as well as \cite{manresa2021assessing}. Nevertheless, comparatively, eradicating gender stereotypes in AI entails enhanced recognition of sociological structures and civilizational norms.

Bias in AI based on gender is not straightforward and has deep layers that mix quantitative–technical and qualitative–social. One type of bias is pseudo-AI Bias, which is in accordance with socially constructed and imposed stereotypes \citep{zhai2022pseudo}. Bias manifests itself in many forms, including skewed views on datasets or bias in the interpretation of results generated by AI systems. This bias is evident within educational situations where the result of an action reflects the performance. It means unequal evaluation, continued perpetration of gender roles, and prescription that keep women off educational attainment. \cite{madaio2022beyond} review the structural bias that could be involved when using AI solutions in learning environments. Lacking precision, scholars will have to develop more elaborate concepts that capture these biases and will have to call for more sophisticated ways of studying them.

The plain truth is that, in education, as in many other areas of life, dispelling myths about gender bias in artificial intelligence, on top of simply being informed about the issue, can go a long way toward building better, fairer systems for all learners. Gendered AI must involve synergy and controversy in ideas of gender studies, practitioners, and theorists in ethical AI and educational theory to ensure that our applications are not just making current dignities inequitable. To tackle AI gender discrimination in education, myths should be disregarded, and the reality is that the actual existence of biases and their emergence is more nuanced within and as a result of the systems of AI. Therefore, such an understanding is necessary to develop AI applications that would be both equitable and efficient in various learning environments.

\subsubsection{This Study's Approach to Bias, Disparities, and Fairness Analysis}
This study employs a rigorous methodology to analyze gender issues in automatic scoring. We have fine-tuned the base variant of the BERT model (similar to \cite{liu2023context}) and GPT-3.5-turbo (identical to \cite{latif2024fine}, using more than 6000 human experts' rated student responses, proportionally distributed among male, female, and mixed datasets across six assessment items. Specific details about the dataset and fine-tuning process are presented in Section 3. This study measures AI gender bias by paired t-test of scoring accuracy between different gender-specifically trained models \citep{mowery2011paired}; the gender disparities by mean score gaps by gender \citep{wang2007automated, wilson2023}; and fairness by Equalized Odds \citep{hardt2016equality}. These methods comprehensively show how gender issues manifest in AI scoring systems for mixed and separately trained models.

Therefore, the concept of the investigation is based on the hypothesis that gender bias in automatic scoring across mixed male and female responses is not apparent when the AI models were trained from mixed-gender datasets. At its center is the idea that it is possible for the discriminative model approaches to learn more gender-neutral patterns with a diverse training set due to the removal of bias from unbalanced datasets. Pertinent to this analysis is its focus on explaining AI and gender issues in a simple manner. Employing a rationale of empirical evidence, the study asserts that when it comes to gender, AI is not averse to running into difficulty but is not necessarily hard-coded. All of them may be solved with proper reasoning and methods during training sessions. The hypothesis rejects the belief that all AI systems are biased or that unbiased AI is too hard to develop.

These results are valuable and relevant for AI advancement in school settings. In conclusion, the authors state that training data must be considered and that applying the right analysis method can lead to more equitable solutions. The conclusions are agnostic to the larger discussion regarding the integration of ethics into the advancement of artificial intelligence theories, stating that all of the theories should serve to teach AIs.

To sum up, this study contributes to the growing literature on gender bias in AI by providing an appropriate course of training and further analysis with improved efficiency. It also brings empirical evidence and conceptual ground, thus opening the path to the possession of apt technical knowledge and thorough moral responsibility to develop decent AI systems devoid of many wrong beliefs.

\section{Methodology}

\begin{figure*}
  \includegraphics[width=\textwidth]{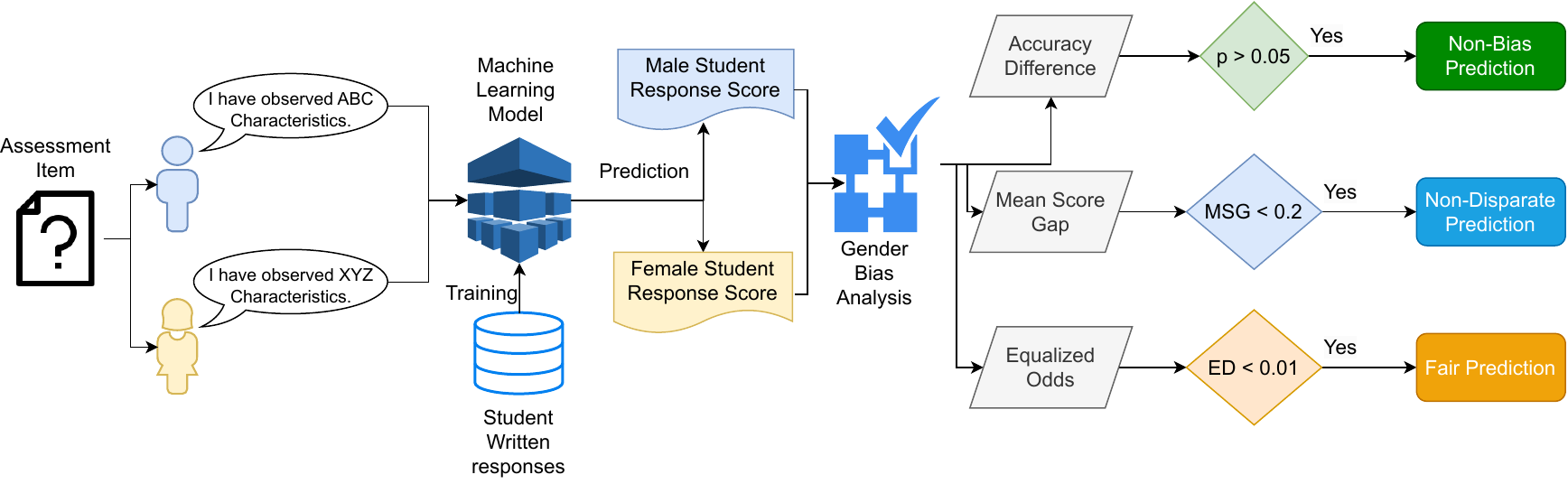}
  \caption{Overview of AI gender bias analysis for automatic scoring.}
  \label{fig:overview}
\end{figure*}

This study investigates gender issues in AI-based scoring systems based on state-of-the-art large language models like BERT and GPT-3.5-turbo models. We have fine-tuned the BERT model on the students' written responses in such a way that the BERT model acts as a multi-class classifier. The human-rated dataset is fed into the BERT model by employing a 3-fold validation mechanism to validate the model's accuracy. Each fold is split into 8:2 for training and validation, and finally, the accumulated accuracy of each fold is reported. Fine-tuning the GPT-3.5-turbo model is a more sophisticated method in which we first process the dataset to clean the data from gibberish letters and extract tokens for model evaluation. GP-3.5-turbo model is initialized by setting ReLU (a regression-based loss function for multi-class classification), setting a small learning rate for better accuracy, and optimal batch size to optimize the fine-tuning process for appropriate epochs. Both models are evaluated on the same matrices consisting of mean absolute error, precision, and accuracy. Large Language models tend to revolutionize education technology \citep{latif2023artificial} and raise potential challenges, such as bias, disparity, and fairness, which must be addressed. A comprehensive methodology was adopted, comprising different training and testing phases for models and employing various statistical techniques to evaluate gender bias, disparities, and fairness. The paired t-test with a p-value $>$ 0.05 on accuracy between machine scores of different genders is used to determine the statistical insignificance and non-biased nature of automatic scoring. The mean score gap with the threshold set to 0.2 determines prediction disparities. Finally, equalized odds are calculated to check fairness in the automatic scoring of written responses by different genders. The threshold for each analysis is extracted by setting parameters for a balanced dataset and scoring accuracy $>$ 80\%. Fig.~\ref{fig:overview} provides a comprehensive overview of the research design and evaluation mechanism. 

\subsection{Dataset}
The study utilized a meticulously categorized dataset of student-written responses to six assessment items, each classified under the multi-class category. The datasets were categorized based on reported gender (i.e., male and female; we eliminated the data that did not report gender information), and a mixed dataset incorporated responses from both genders. The dataset was collected from the Mathematical Thinking in Science (MTS) project \citep{jin2019validation, ETS2023}. Based on the availability of data, we used the maximum of available data while maintaining the ratio between training and testing data for each item by gender. Consequently, we randomly identified three training datasets for each item: mixed, male, and female training datasets and three testing datasets for each item. The detailed composition of each assessment item's dataset is presented in Table~\ref{table:datasets}. This comprehensive dataset facilitated a nuanced analysis of gender bias in AI-based scoring systems, ensuring robust and broadly applicable study findings.

\begin{table*}[htp]
\centering
\caption{Summary of Datasets and Their Samples}
\label{table:datasets}
\begin{tabular}{|p{2cm}|p{1.5cm}|p{1.5cm}|p{1.5cm}|p{1.5cm}|p{1.5cm}|p{1.5cm}|}
\hline
\textbf{Datasets}  & \textbf{mixed Training samples} & \textbf{mixed Testing samples} & \textbf{Male Training samples} & \textbf{Male Testing samples} & \textbf{Female Training samples} & \textbf{Female Testing samples} \\ \hline
falling weights          & 1148                                & 230                               & 342                        & 87                         & 478                        & 120                       \\ \hline
gelatin                  & 918                                 & 230                               & 344                        & 86                         & 477                        & 120                       \\ \hline
bathtub                  & 916                                 & 230                               & 364                        & 92                         & 450                        & 113                       \\ \hline
sandwater1               & 915                                 & 229                               & 364                        & 92                         & 450                        & 113                       \\ \hline
sandwater2               & 913                                 & 229                               & 364                        & 92                         & 449                        & 113                       \\ \hline
two boiling situations    & 909                                 & 228                               & 363                        & 91                         & 448                        & 113                       \\ \hline
\end{tabular}
\end{table*}

\subsubsection{Dataset Description:}
In the multi-class assessment activities, students must use mathematical reasoning (such as proportional reasoning) and scientific knowledge to answer science problems (Jin et al., 2019). The purpose of the tasks is to evaluate the mathematical reasoning in science and the process by which students learn it as they advance through science courses in high school. The high school-level tasks focus on two NGSS disciplinary fundamental ideas: PS3 (Energy) and LS2 (Ecosystems: Interactions, Energy, and Dynamics). Almost 6000 students in grades 9 through 12 participated in the study. Eight educators scored the students' answers using the respective scoring rubrics. To respond to these tasks, students should consider the three key concepts: 1) The two variables that are dependent on the amount of material (mass, M), heat (Q), and temperature (T). Heat is an extensive variable, whereas temperature is an intense variable independent of the amount of substance. 2) Specific heat as a compound is needed to increase a substance's temperature by one degree Celsius per unit of mass. 3) T depends on M and C for an identical quantity of Q: The intake and outflow of heat will change a substance's temperature.

\begin{figure}[hbt!]
    \centering
    \includegraphics[width=1\linewidth]{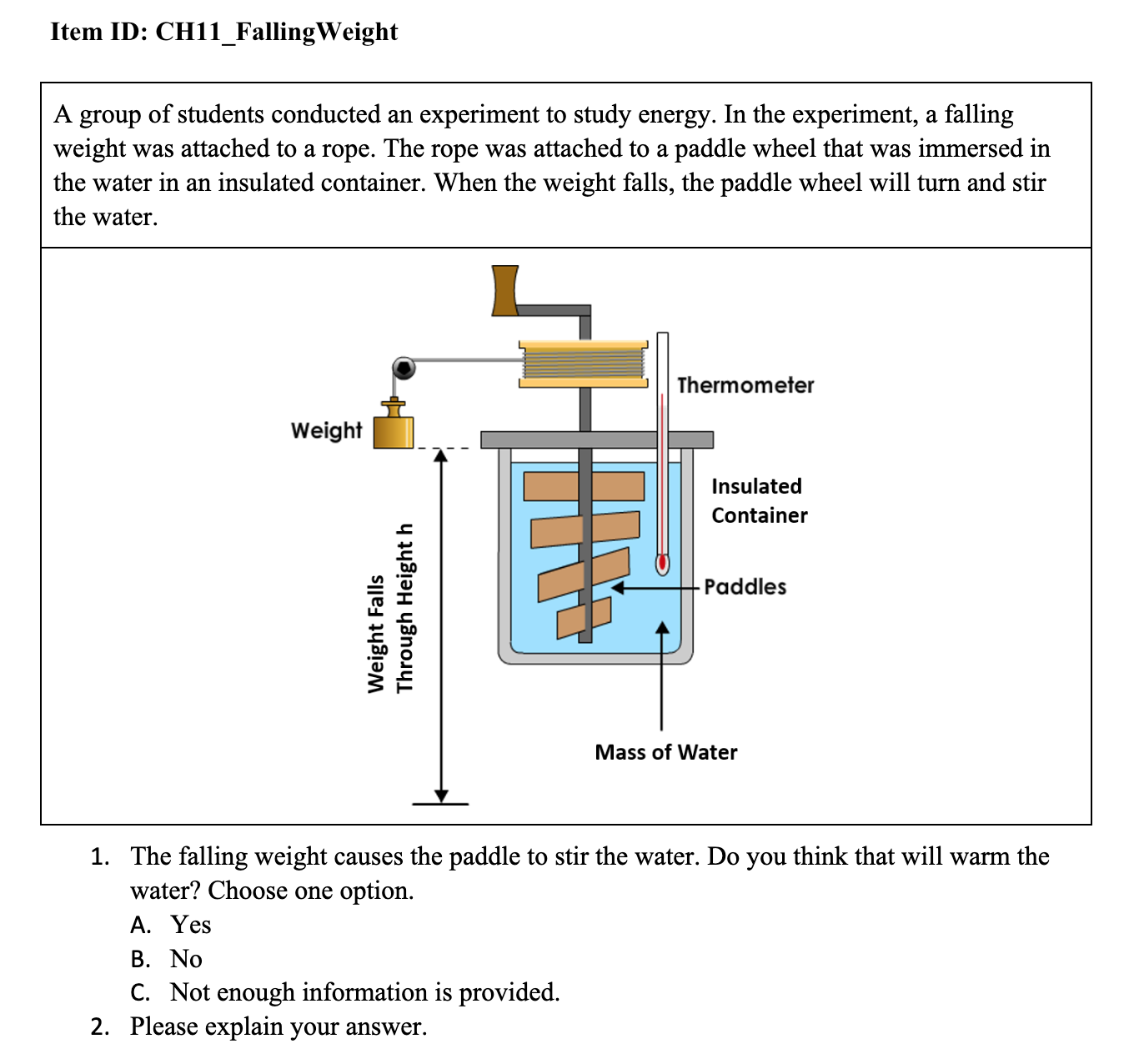}
    \caption{Example Multi-class Task: Falling weights}
    \label{fig:falling_weights}
\end{figure}

Fig.~\ref{fig:falling_weights} presents an example of multi-class tasks that illustrate a situation where a paddle wheel is driven by a falling weight to agitate the water. Students must predict whether the water will warm up and provide a justification in writing. As a result of the paddle stirring the water due to falling weight, students must determine the heat change in the water in this example problem. A four-level rubric was created to evaluate students' knowledge usage proficiency in answering the example item. Human experts from ETS have graded the student responses based on the rubric. Specific details for each level are listed in Table~\ref{table:rubric_falling_weights}.

\begin{table} [hbt!]
\centering
\caption{Scoring rubric for task: Falling weights}
\begin{tabular}{l p{5.5cm} p{5.5cm} }
\toprule
\textbf{Level} & \textbf{Description} & \textbf{Example} \\
\midrule
3 & A Level 3 response suggests that the student understands the measurability of variables. Level 3 responses must choose A. There are two patterns at Level 3: & Example 1: A [Yes.] Because the weight will stir the water, the movement in water particles will cause the temperature to rise. \\
& \textbf{3a:} [Identification of heat/energy] The response 1) associates movement (e.g., movement of the weight, paddle, and/or water) with energy/heat AND/OR 2) associates particle movement with temperature, heat, or energy. & Example 2: A. [Yes.] The falling weight transfers its energy to the paddles that spin. The spinning paddles transfer their energy to the water. Because water (the system) absorbs energy, the temperature of the water will increase even if it is very small. \\
& \textbf{3b:} [Identification of heat/energy with errors.] The response 1) associates movement (weight, paddle, and/or water) with energy or heat AND/OR 2) associates particle movement with temperature, heat, or energy. But, the response also confuses energy/heat/temperature with other variables such as forces. & Example 3: A. [Yes.] The weight falling will cause the paddle to start moving, which turns gravity into kinetic energy, and the kinetic energy is then transferred to the water and moves the water's molecules, which would then start heating the water. \\

2 & A Level 2 response: & Example 1: C [Not enough information is provided.] It depends on how fast the paddles are turning. If they are turning fast enough, they could warm the water.\\
& \textbf{2a}: [Threshold] The response indicates a threshold where the movement must be fast to a certain degree or the energy input must be large enough to cause an increase in temperature. & Example 2: B. [No.] Stirring the water will not increase the temperature of the water because no energy is being added. \\
& \textbf{2b:} [General understanding] The response indicates that the student generally understands that work/energy will cause the temperature to increase but cannot apply that general understanding to this specific scenario. & Example 3. C. [Not enough information is provided.] there is not enough information to explain more about the temperature; just will it spin, and yes, it will spin pretty fast. \\
& \textbf{2c:} [Macroscopic causation] The response provides a causal relationship at a macroscopic scale without using energy or heat. & Example 4: B. [No.] We don't know if the water is warm or cold already, room temperature, etc.\\
& \textbf{2d:} [Irrelevant variables ONLY] The response analyzes the scenario based on irrelevant variables. & Example 5: A. [Yes] Due to the friction between the paddle and the water, the water will get warmer after some time.\\
1 & Level 1 responses have the following patterns: & Example 1: A. [Yes.] It's placed in an insulated container, and it has a thermometer. \\
& \textbf{1a:} [IDK] "I don’t know" type of response. &  \\
& \textbf{1b:} [No information] The response does not provide information about the student’s ideas about the phenomenon or data. &  \\
0 & Blank or random letters & Acareba artouth \\
\bottomrule
\end{tabular}
\label{table:rubric_falling_weights}
\end{table}

\subsection{Experimental Setup}

The experimental setup involved rigorous training and testing using multiple training and testing datasets. 

\textit{Training Phase:} Each dataset undergoes a unique training cycle:
\begin{itemize}
    \item A mixed-trained model trained with the mixed training dataset including both genders.
    \item A male-trained model trained exclusively with data from males.
    \item A female-trained model trained solely with data from females.
\end{itemize}
Using this approach, we developed three distinct language models for each assessment item, fine-tuned using both BERT and GPT-3.5, respectively.

\textit{Testing Phase:} Every trained model goes through several testing stages to determine possible bias. Specifically,
In the first experiment, the model trained on the mixed dataset was compared to the mixed testing data as well as separately for the male and female testing data. We trained and utilized both BERT and GPT-3.5. In the second experiment, Gender-specific models were used to test against the same and the counterpart's gender datasets. Likewise, we employed BERT and GPT-3.5. This experimental design was intended to systematically evaluate any possible prejudices in the models' decision-making based on datasets and gender splits.

\subsection{Accuracy Difference (Pair t-Test)}

Paired t-tests were employed to analyze the difference in accuracy between trained models of related. \cite{mowery2011paired} noted that the paired t-test is useful in establishing the differences in pairs of observations. In our work, it was imperative to compare the accuracies of models that were trained using different datasets.

Mathematically, the paired t-test is expressed as:
\begin{equation}
    t = \frac{\overline{D}}{s_D / \sqrt{n}}
\end{equation}
where $\overline{D}$ is the mean of the differences between paired observations, $s_D$ is the standard deviation of the differences, and $n$ is the number of pairs. A p-value greater than 0.05 indicated no significant difference in accuracies between the models, suggesting minimal gender bias.

\subsection{Mean Score Gap}

Mean Score Gap (MSG) was utilized to evaluate the disparity in scores between different genders. MSG can be generated either by human experts or machine algorithmic models. Given that human-assigned scores have undergone rigorous scrutiny, we used MSG generated by human-assigned scores as the golden standard to evaluate machine-generated MSG. It can be seen that if machine-generated MSG is larger than human-generated MSG for the same testing data, the machine exacerbates the gender difference and may potentially create inequity. \cite{wang2007automated} and \cite{wilson2023} highlighted the importance of such comparisons in their study on automated essay scoring and automatic scoring of written scientific arguments. MSG for target dataset $\mathbf{K}$ is calculated as:
\begin{equation}
    MSG_\mathbf{K} = \frac{\sum_{i=1}^{n} score_{\mathbf{K},i}}{n} - \frac{\sum_{i=1}^{m} score_{\mathbf{K},i}}{m}
\end{equation}
where $score_{k,i}$ is the score assigned to the $i^{th}$ response by models trained on dataset $\mathbf{K}$; where $\mathbf{K} = \{K_{male}, K_{female}, K_{mix} \}$, and $n$, $m$ are total number of male and female responses in testing data. A threshold of MSG $<$ 0.2 was used to indicate acceptable score disparity.

\subsection{Equalized Odds}

Equalized Odds, as defined by Hardt et al. \cite{hardt2016equality}, involves assessing the equality of true and false positive rates across groups. It is a crucial measure of fairness in predictive models. The Equalized Odds (EO) criterion is given by:
\begin{equation}
    EO = max(|TPR_{male} - TPR_{female}|, |FPR_{male} - FPR_{female}|)
\end{equation}
where $TPR_{male}$ and $TPR_{female}$ are the true positive rates for Male and Female, respectively, and $FPR_{male}$ and $FPR_{female}$ are the false positive rates for the two genders. An EO value less than 0.01 indicates a fair model with minimal gender bias.

These three methods - Paired t-test, MSG, and Equalized Odds - provide a comprehensive approach to evaluating gender bias, disparities, and fairness in AI-based scoring systems, ensuring the ethical uses of AI models. The methodology encompassed a detailed experimental setup with rigorous training and testing phases, coupled with statistical analyses to comprehensively assess gender issues in AI-based scoring systems.

\section{Results}

\subsection{Scoring Accuracy Difference Evaluation}
A paired samples t-test was performed to evaluate whether there was a statistical difference between the scoring accuracy for male and female predicted scores before and after randomly mixing both genders' responses.

A repeated measures ANOVA was conducted to compare the scoring accuracy of mixed testing data from the mixed-trained BERT model and GPT-3.5 model with their respective male-trained and female-trained counterparts.

For the BERT model, the analysis revealed that the scoring accuracy of mixed testing data from the mixed-trained model ($\Delta M = 0.022$, $SD = 0.022$) was not significantly higher than that of the male-trained model ($\Delta M = 0.020$, $SD = 0.080$), $t(6) = -0.86$, $p = .42$. Similarly, no significant difference was observed between the mixed-trained model and the female-trained model ($\Delta M = 0.030$, $SD = 0.069$), $t(6) = -1.37$, $p = .22$.

For the GPT-3.5 model, the scoring accuracy of mixed testing data from the mixed-trained model ($\Delta M = 0.023$, $SD = 0.090$) was also not significantly higher than that of the male-trained model ($\Delta M = 0.016$, $SD = 0.110$), $t(6) = -0.42$, $p = .53$, or the female-trained model ($\Delta M = 0.018$, $SD = 0.010$), $t(6) = -0.42$, $p = .69$.

Both the BERT and GPT-3.5 models demonstrated consistent performance across mixed and gender-specific datasets, suggesting minimal gender biases. 

\begin{figure*}[ht]
 \includegraphics[width=0.32\linewidth]{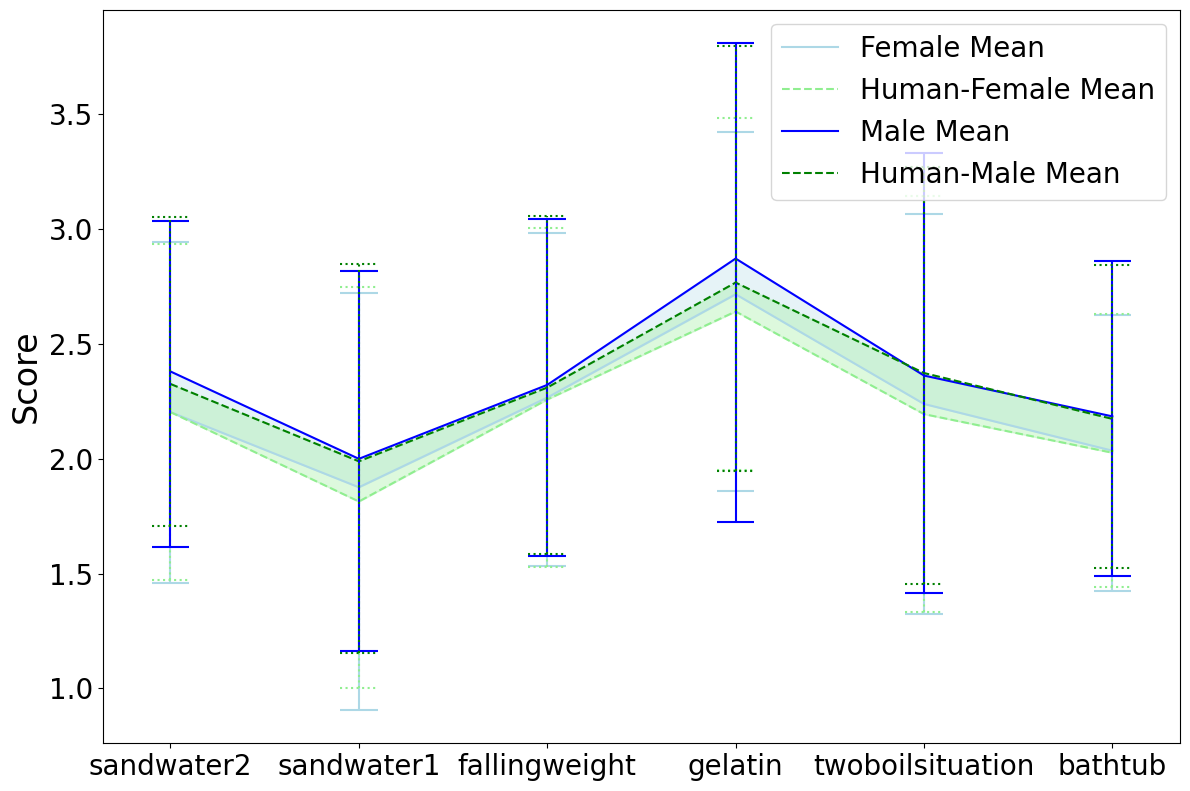}
 \includegraphics[width=0.32\linewidth]{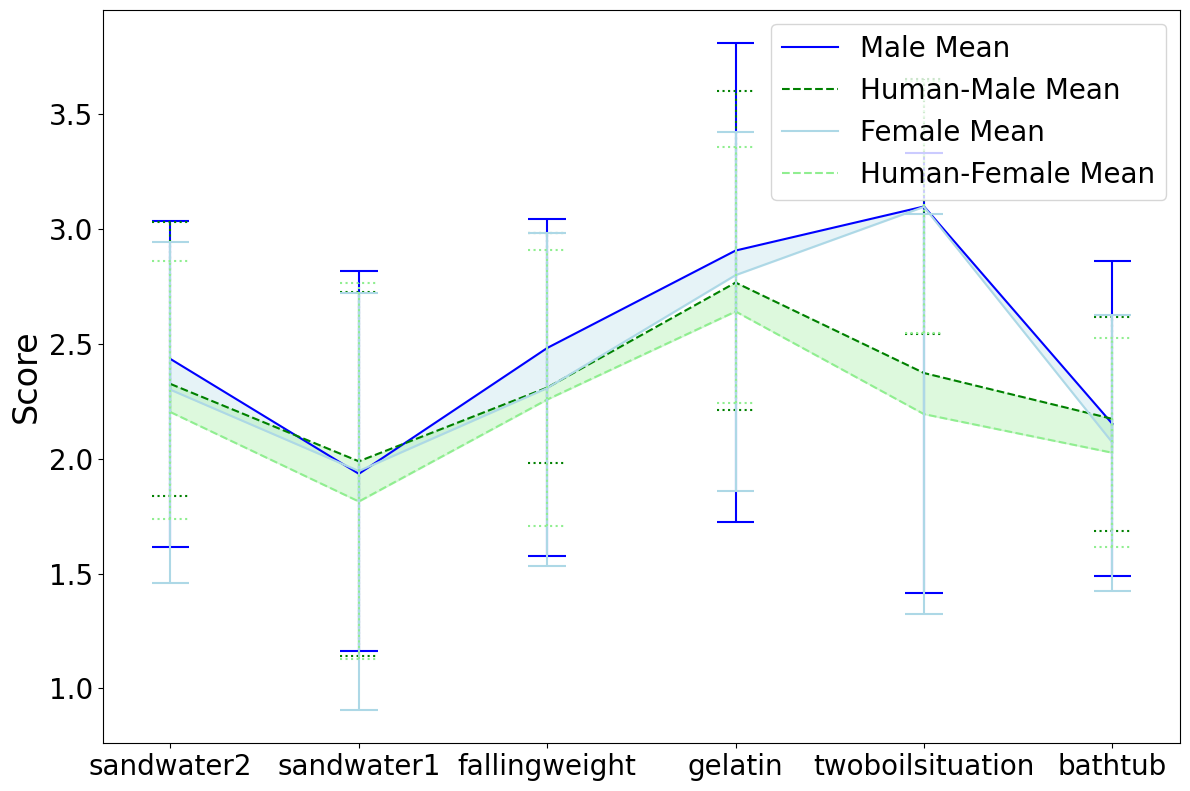}
 \includegraphics[width=0.32\linewidth]{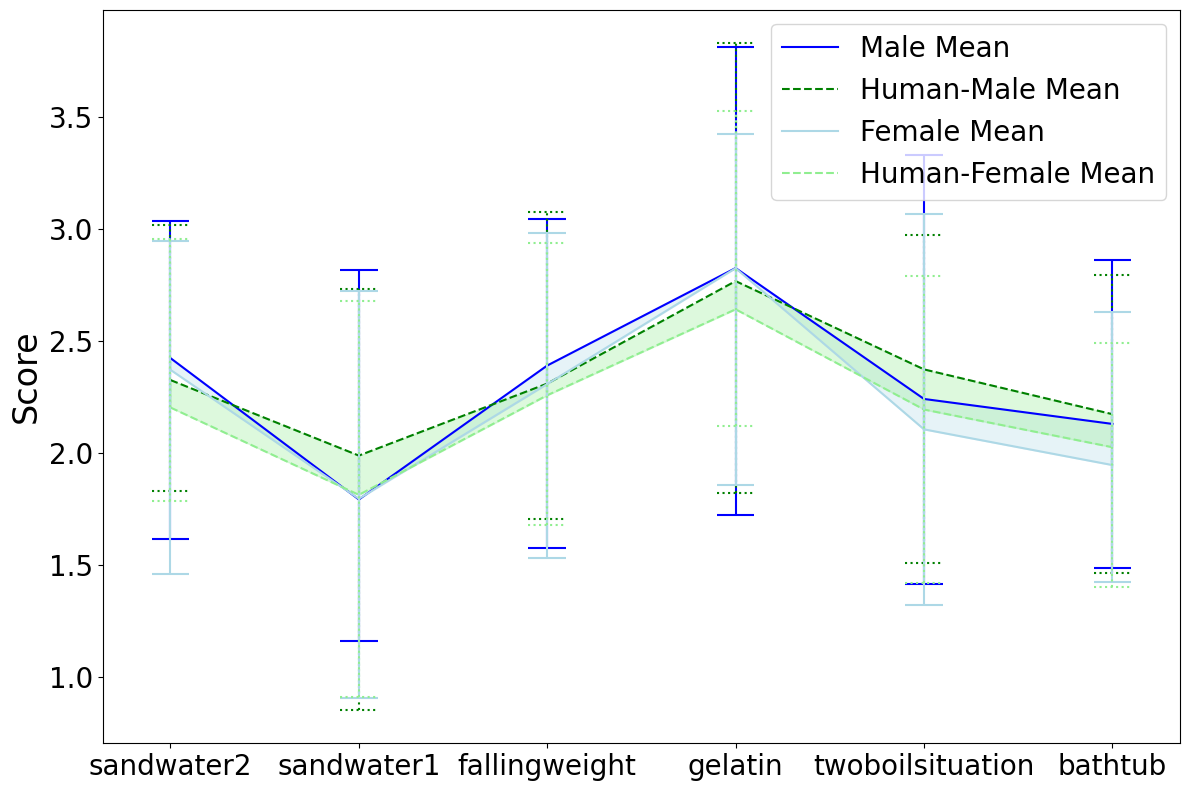}
\includegraphics[width=0.32\linewidth]{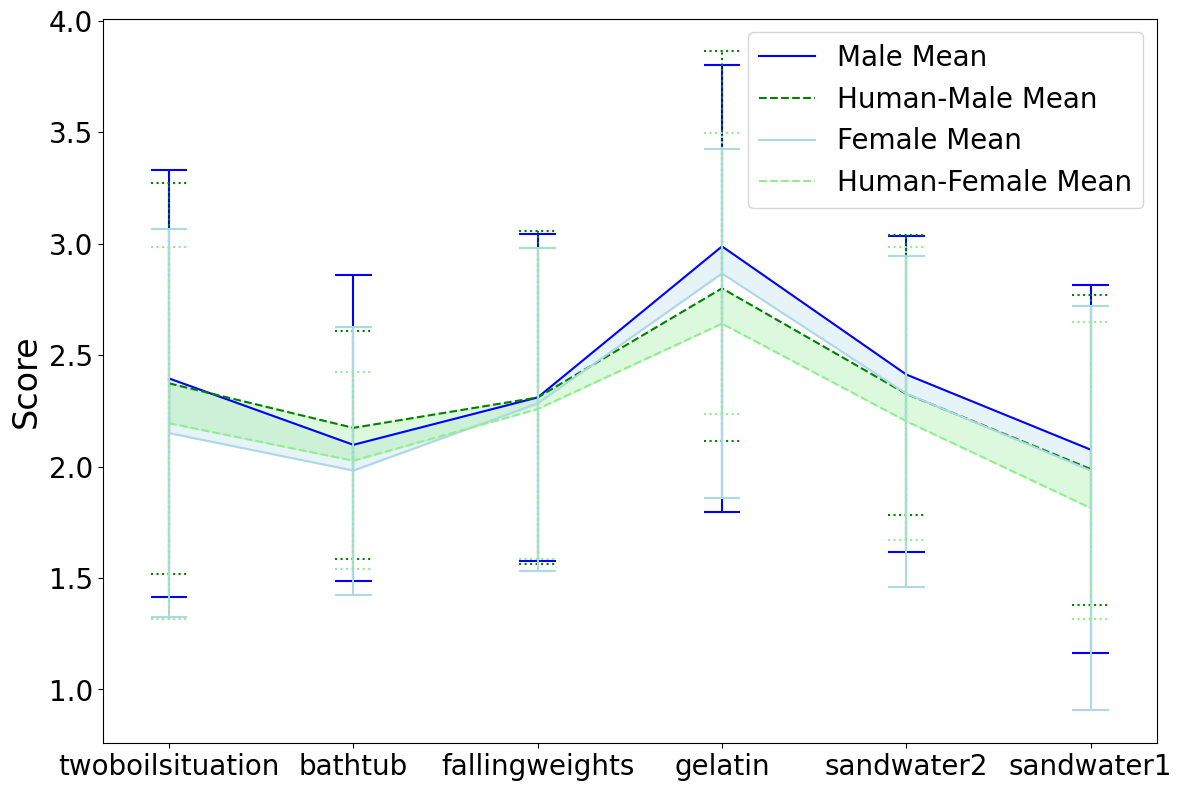}
 \includegraphics[width=0.32\linewidth]{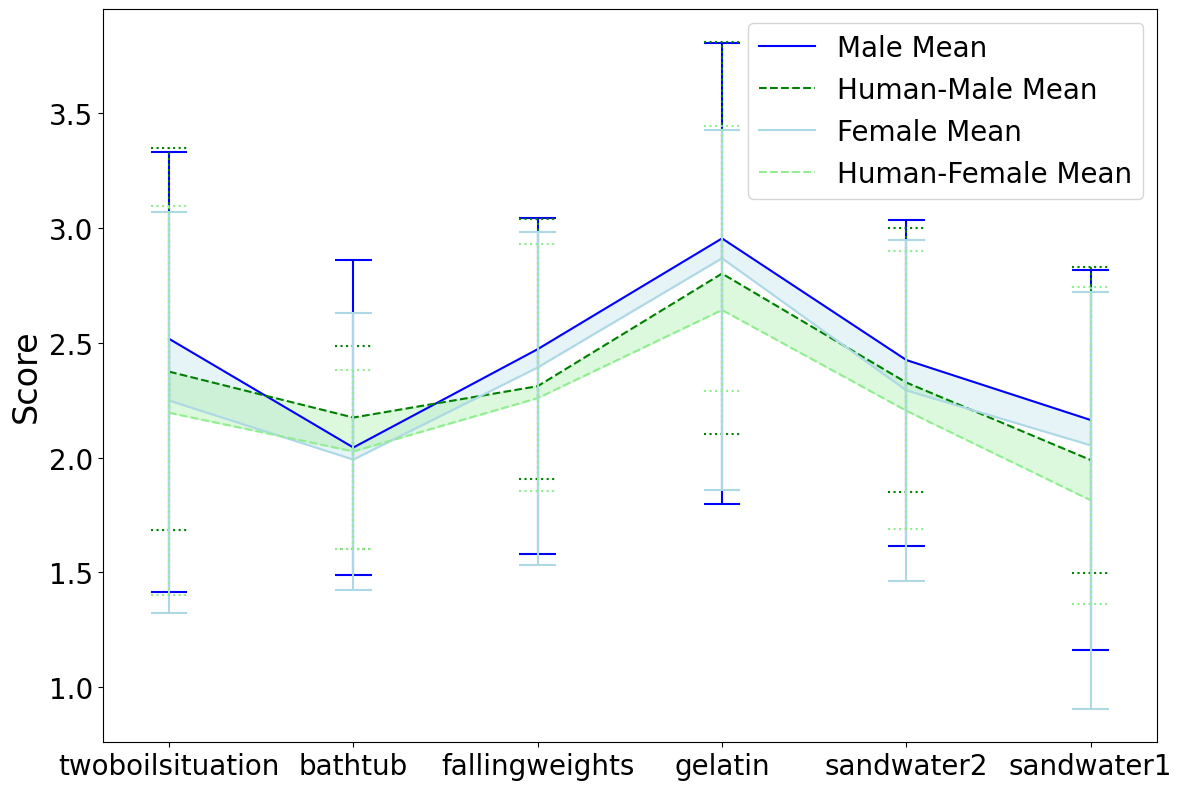}
\includegraphics[width=0.32\linewidth]{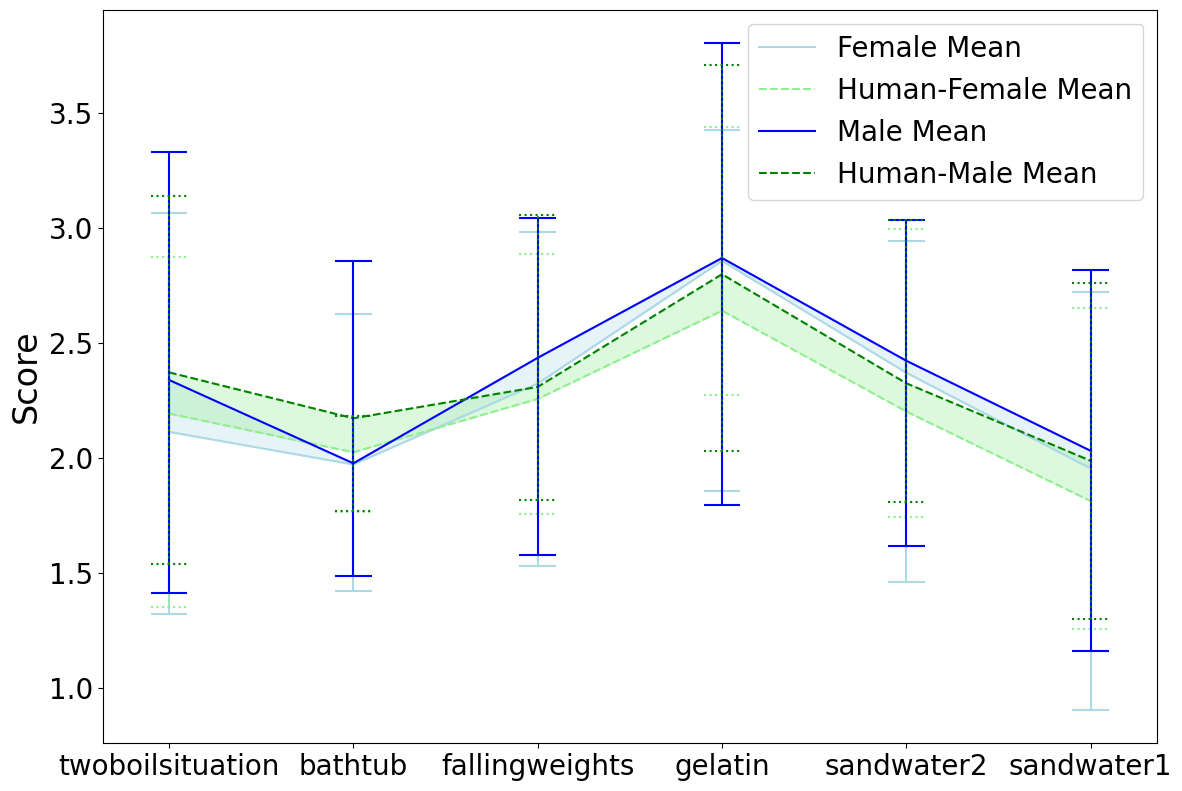}
\caption{Mean score gaps between male and female testing data: Comparing human-graded scores and fine-tuned LLMs' scores: Mixed training Model \textbf{(Left)}, Male trained Model \textbf{(Center)}, and Female trained Model \textbf{(Right)} comparison line plots of BERT \textbf{(Top)} and GPT3.5 \textbf{(Bottom)} fine-tuned models's scores. The shaded region in each plot signifies the $\Delta MSG$.}
 \label{fig:comparison}
\end{figure*}

\subsection{Mean Score Gap}

To verify our assumption, we designed three experiments to compare the difference of MSGs ($\Delta MSG$) between human- and machine-graded scores. For each experiment, we tested the three models: Mixed-, male-, and female-trained models and compared each of them with human-generated MSG on combined male and female testing samples.

A repeated measures ANOVA was conducted to analyze the $\Delta MSG$ between human and mixed-trained machine-graded scores for BERT and GPT-3.5 models. This was to validate the hypothesis that mixed-gender training data would have the least impact on the MSG. Human scores, used as the golden standard, yielded an average MSG of $M_{\text{Human}} = -0.0957$, $SD_{\text{Human}} = 0.0746$. 

In the first experiemnt, upon training the mixed-gender model, the BERT model produced an average MSG of $M_{\text{Mix-BERT}} = -0.1078$, $SD_{\text{Mix-BERT}} = 0.0742$, with a difference of $\Delta MSG_{\text{Mix-BERT}} = 0.0122$ compare to $M_{\text{Human}}$. Similarly, the GPT-3.5 model yielded an average MSG of $M_{\text{Mix-GPT}} = -0.1150$, $SD_{\text{Mix-GPT}} = 0.0724$, with the MSG difference of $\Delta MSG_{\text{Mix-GPT}} = 0.0193$ compare to $M_{\text{Human}}$. However, these differences were not statistically significant for both BERT and GPT mixed models with values ($t_{\text{Mix-BERT}} = -1.5598$, $p > 0.05$) and ($t_{\text{Mix-GPT}} = -1.8896$, $p > 0.05$), respectively.

In the second experiment, for the gender-specific models, the BERT model demonstrated a decrease in the average MSGs compared to human scores: $M_{\text{Male, BERT}} = -0.0742$, $SD_{\text{Male, BERT}} = 0.0871$, and $M_{\text{Female, BERT}} = -0.0603$, $SD_{\text{Female, BERT}} = 0.0564$. The MSG differences were $\Delta MSG_{\text{Male, BERT}} = -0.0337$ and $\Delta MSG_{\text{Female, BERT}} = -0.0138$. Both male and female BERT models have shown a reduction in visible MSG in their correlations $\Delta MSG$ compared to human MSG and have a statistically significant difference ($t _{\text{Male, BERT}} = -3.0857$, $p < 0.05$) and ($t _{\text{Female, BERT}} = -3.6226$, $p < 0.05$).

Similarly, for the GPT-3.5 model, both gender-specific models demonstrated a lower average MSG ($M_{\text{Male, GPT}} = -0.1212$, $SD_{\text{Male, GPT}} = 0.0769$) compared to the female-specific model ($M_{\text{Female, GPT}} = -0.0808$, $SD_{\text{Female, GPT}} = 0.0815$). The MSG differences were $\Delta MSG_{\text{Male, GPT}} = 0.0062$ and $\Delta MSG_{\text{Female, GPT}} = -0.043$, with visible statistically significant differences ($t _{\text{Male, GPT}} = -3.8574$, $p < 0.05$) and ($t _{\text{Female, GPT}} = -3.4283$, $p < 0.05$). (Detailed results can be seen in Fig.~\ref{fig:comparison}).

These results indicate that training with a mixed dataset in both BERT and GPT-3.5 models has shown a minimal MSG compared to gender-specific training. This outcome validates our hypothesis that a balanced mixed training dataset has no significant traces toward gender disparities, as reflected in the similar MSGs in the mixed models compare to human MSG. On the other hand, gender specific models has shown siginifacant tilt toward gender disparities. The findings underscore the importance of inclusive training datasets in developing AI models that yield more equitable scoring outcomes.

\subsection{Equalized Odds Evaluation}

The study evaluated the Equalized Odds (EO) \citep{hardt2016equality} for BERT and GPT-3.5 models to assess fairness in gender predictions. 

The BERT model demonstrated an EO of 0.042 on the mixed testing dataset for the mixed-trained model, indicating a lower disparity in prediction fairness. In contrast, the gender-specifically-trained models exhibited higher EO values on the mixed testing dataset, with the Male model showing an EO of 0.107 and the Female model having an EO of 0.074. These results suggest that the mixed-trained model is more equitable in its predictions across genders than gender-specific models.

For the GPT-3.5 model, the mixed-trained model's EO was 0.061 on the mixed testing dataset, again reflecting a reduced gradient and, thus, greater fairness in predictions. The Male and Female trained models showed EOs on the same testing dataset of 0.076 and 0.074, respectively. While these values are closer to that of the mixed model than the BERT model, they indicate a higher disparity in the gender-specific models.

The mixed-trained models for both BERT and GPT-3.5 have lower EO values, suggesting more equitable predictions and higher fairness than gender-specific models. This outcome supports the hypothesis that training with a balanced dataset can reduce the EO gradient, contributing to the development of fair AI models.

\section{Discussion}
The intelligent computational analysis plays a role when it comes to evaluating the student's performance and defining existing learning capacity. Considering a great deal of data, AI systems may show the teachers information and the potentiality of how the students use them, thus giving them a better perspective of decision-making and enclosing better strategies. Analyzing the impact of gender bias, disparities, and fairness when it comes to the various aspects of the automatic scoring issues and the gender biases of the data sets that have been used, the presented research framework incorporates an experimental method. In a study where we evaluated the efficacy of gender-unbalanced training datasets in the elimination of gender bias in the scored AI system, it was established to be useful. Combined with this, it also introduced facts that endeavored to provide real evidence of sources in excluding the prejudice that AI only escalates sexism \citep{bolukbasi2016man, lu2020gender}.

On the contrary, this study found that algorithms for imbalanced data may further reinforce gender-based bias instead of addressing this issue. In this study, the balanced mixed training model did not exhibit gender-biased behaviors, so this study offered an opportunity to address such baseless claims about the prejudice of AI systems. To avoid acting as the malevolent agent that deepens the impression of gender-predetermined roles and functions that are assigned to women, non-transphobic gender minorities' representations in the data that feeds models must be addressed.

As for the concept of balanced training for female and male models, the results obtained based on the proposed mixture training are provided by the various minimized EO gradients and decreased MSG in the mixture-trained models. This discovery is crucial given that \cite{hardt2016equality}. The authors have further affirmed the role of opportunity equality in supervised learning. Based on the findings of our study, it can be concluded that mixed training datasets with all the data split in fifty-fifty could enhance the fairness of the AI-based tests and assessment in the learning context where the fairness of the AI is paramount \citep{li2022using}. This discovery establishes the relevance and fairness of AI functions in higher learning. It falls in line with the fundamental postulations presented by \cite{slimi2023navigating} with regard to the use of AI.

Our findings align with the literature supporting debiasing AI systems \citep{manresa2021assessing, nadeem2022gender}. This work shows that balanced training produces more equitable AI outcomes, which advances the more significant attempts to develop impartial and fair AI models. The method offers a workable answer to this widespread problem by refuting that AI systems are bound to reproduce societal prejudices. The study's findings strongly support the gender issue of balanced mixed training in AI for automated scoring. In addition to addressing the pressing issues of equality and fairness in AI-based scoring systems, our method offers proof against mechanical pseudo-bias in general AI applications.

\section{Conclusion}
The analysis using BERT and GPT-3.5 on student written responses has significantly advanced our understanding of gender bias, disparities, and fairness in AI-based scoring systems. The paper also shows the concerns of gender imbalance training, which has nearly no gender bias. In this regard, the results are, in a way, opposite to what one would expect given the current discourses about the biases that AI may inherently possess. This meant that the models trained in mixed genders, rather than in the single-gender data set, had a lower MSG than humans; it was argued that algorithms trained in the imbalanced gender data set could help expand gender bias. It was seen from the analysis employing EO that gender intermixed underwent far less bias than gender-exclusive models. The research findings reveal that gender-sparse data does not per se lead to scoring bias but amplifies gender inequalities and overall scoring equity.

\section*{Acknowledgment}
This study secondary analyzed data from a project supported by the Institute of Education Sciences (grant number
R305A160219). The authors acknowledge the funding agencies and the project teams for making the data available for analysis. Project is partially supported by NSF (grants numbers 2101104). The findings, conclusions, or opinions herein represent the views of the authors and do not necessarily represent the views of personnel affiliated with the funding agencies.


\bibliography{sn-bibliography}

\end{document}